\documentclass[12pt, onecolumn]{IEEEtran}
\usepackage{amsmath}
\usepackage{amssymb}
\usepackage{mathrsfs}
\usepackage{cite}
\usepackage{epsfig}
\usepackage{epsf}
\usepackage{theorem}
\usepackage{graphics}
\usepackage[active]{srcltx}

\begin{document}

\title{Green Modulations in Energy-Constrained \\
Wireless Sensor Networks \\
\thanks{The work was supported in
part by an Ontario Research Fund (ORF) project entitled
``Self-Powered Sensor Networks''. The work of Jamshid Abouei
was performed when he was with the Dept. of ECE, University of
Toronto, Toronto, ON M5S 3G4, Canada. The material in this paper was
presented in part to ICASSP'2010 conference, March 2010
\cite{Jamshid_ICASSP2010}.}}

\author{\small  Jamshid Abouei$^{\dag}$, Konstantinos N.
Plataniotis$^{\dag \dag}$, and Subbarayan Pasupathy$^{\dag \dag}$\\
$^{\dag}$ Department of Electrical Engineering, Yazd University, Yazd, Iran, Email:
abouei@yazduni.ac.ir\\
$^{\dag \dag}$ The Edward S. Rogers Sr. Dept. of ECE, University of Toronto, Toronto, ON M5S 3G4, Canada\\
Emails: \{kostas, pas\}@comm.utoronto.ca}

\maketitle

\begin{abstract}
Due to the unique characteristics of sensor devices, finding the
energy-efficient modulation with a low-complexity implementation
(refereed to as \emph{green modulation}) poses significant
challenges in the physical layer design of Wireless Sensor Networks
(WSNs). Toward this goal, we present an in-depth analysis on the
energy efficiency of various modulation schemes using realistic
models in the IEEE 802.15.4 standard to find the optimum
distance-based scheme in a WSN over Rayleigh and Rician fading
channels with path-loss. We describe a proactive system model
according to a flexible duty-cycling mechanism utilized in practical
sensor apparatus. The present analysis includes the effect of the
channel bandwidth and the active mode duration on the energy
consumption of popular modulation designs. Path-loss exponent and
DC-DC converter efficiency are also taken into consideration. In
considering the energy efficiency and complexity, it is demonstrated
that among various sinusoidal carrier-based modulations, the
optimized Non-Coherent M-ary Frequency Shift Keying (NC-MFSK) is the
most energy-efficient scheme in sparse WSNs for each value of the
path-loss exponent, where the optimization is performed over the
modulation parameters. In addition, we show that the On-Off Keying
(OOK) displays a significant energy saving as compared to the
optimized NC-MFSK in dense WSNs with small values of path-loss
exponent.
\end{abstract}

\begin{center}
\end{center}

\section{Introduction}
Wireless Sensor Networks (WSNs) have been recognized as a collection
of distributed nodes to support a broad range of applications,
including monitoring, health-care and detection of environmental
pollution. In such configuration, sensors are typically powered by
limited-lifetime batteries which are hard to be replaced or
recharged. On the other hand, since a large number of sensors are
deployed over a region, the circuit energy consumption is comparable
to the transmission energy due to the short distance between nodes.
Thus, minimizing the total energy consumption in both circuits and
signal transmission is a crucial task in designing a WSN \cite{Jamshid_ICASSP2010, Jamshid_QBSC2010}. Central to
this study is to find energy-efficient modulations in the physical
layer of a WSN to prolong the sensor lifetime. For this purpose,
energy-efficient modulations should be simple enough to be
implemented by state-of-the-art low-power technologies, but still
robust enough to provide the desired service. In addition, since
sensor devices frequently switch from sleep mode to active mode,
modulator circuits should have fast start-up times. We refer to
these simple and low-energy consumption schemes as \emph{green
modulations}.

In recent years, several energy-efficient modulations have been
studied in the physical layer of WSNs (e.g., \cite{Qu_ITSP0908,
GarzasEURASIP2007}). In \cite{Qu_ITSP0908} the authors compare the
battery power efficiency of PPM and OOK based on the Bit Error Rate
(BER) and the cutoff rate of a WSN over path-loss Additive White
Gaussian Noise (AWGN) channels. Reference \cite{GarzasEURASIP2007}
investigates the energy efficiency of a centralized WSN with an
adaptive MQAM scheme. However, adaptive approaches impose some
additional system complexity due to the multi-level modulation
formats plus the channel state information fed back from the sink
node to the sensor node. Most of the pioneering work on
energy-efficient modulations, including research in
\cite{Qu_ITSP0908}, has focused only on minimizing the average
energy consumption per information bit, ignoring the effect of the
bandwidth and transmission time duration. In a practical WSN,
however, it is shown that minimizing the total energy consumption
depends strongly on the active mode duration and the channel
bandwidth \cite{Cui_GoldsmithITWC0905}.

In this paper, we present an in-depth analysis (supported by
numerical results) of the energy efficiency of various modulation
schemes considering the effect of the ``\emph{channel bandwidth}''
and the ``\emph{active mode duration}'' to find the distance-based
green modulations in a proactive WSN. For this purpose, we describe
the system model according to a flexible duty-cycling process
utilized in practical sensor devices. This model distinguishes our
approach from existing alternatives \cite{Qu_ITSP0908}. New analysis
results for comparative evaluation of popular modulation designs are
introduced according to the realistic parameters in the IEEE
802.15.4 standard \cite{IEEE_802_15_4_2006}. We start the analysis
based on a Rayleigh flat-fading channel with path-loss which is a
feasible model in static WSNs \cite{Qu_ITSP0908}. Then, we evaluate
numerically the energy efficiency of sinusoidal carrier-based
modulations operating over the more general Rician model which
includes a strong direct Line-Of-Sight (LOS) path. Path-loss
exponent and DC-DC converter efficiency (in a non-ideal battery
model) are also taken into consideration. It is demonstrated that
among various sinusoidal carrier-based modulations, the optimized
Non-Coherent M-ary Frequency Shift Keying (NC-MFSK) is the most
energy-efficient scheme in sparse WSNs for each value of the
path-loss exponent, where the optimization is performed over the
modulation parameters. In addition, we show that the On-Off Keying
(OOK) has significant energy saving as compared to the optimized
NC-MFSK in dense WSNs with small values of path-loss exponent.
NC-MFSK and OOK have the advantage of less complexity and cost in
implementation than MQAM and Offset-QPSK used in the IEEE 802.15.4
protocol, and can be considered as green modulations in WSN
applications.

The rest of the paper is organized as follows. In Section
\ref{model_Ch2}, the proactive system model and assumptions are
described. A comprehensive analysis of the energy efficiency for
popular sinusoidal carrier-based and UWB modulations is presented in
Sections \ref{Analysis_Ch 3} and \ref{Analysis_Ch 4}. Section
\ref{simulation_Ch5} provides some numerical evaluations using
realistic models to confirm our analysis. Finally in Section
\ref{conclusion_Ch6}, an overview of the results and conclusions are
presented.

For convenience, we provide a list of key mathematical symbols used
in this paper in Table I.

\begin{table}
   \label{table234}
\caption{List of Notations } \centering
  \begin{tabular}{|l|l|}
  \hline
  $B$:~Bandwidth                          & $b$:~Number of bits per symbol                          \\
  $B_{eff}$:~Bandwidth efficiency         & $\mathcal{E}_t$:~Energy per symbol                      \\
  $d$:~Distance                           & $\mathcal{E}_{N}$:~Total energy consumption for $N$-bit \\
  $M$:~Constellation size                 & $h_{i}$:~Fading channel coefficient                     \\
  $P_s$:~Symbol error rate                & $\mathcal{L}_d$:~Channel gain factor in distance $d$    \\
  $T_s$:~Symbol duration                  & $N$:~Number of transmitted bits in active mode period   \\
  $T_{ac}$:~Active mode period            & $\mathcal{P}_c$:~Total circuit power consumption        \\
  $T_{tr}$:~Transient mode period         & $\mathcal{P}_t$:~RF transmit power consumption per symbol\\
  $\eta$:~Path-loss exponent              & $\chi_e$:~Power transfer efficiency in DC-DC converter  \\
  $\Omega=\mathbb{E}\left[\vert h_{i} \vert^2\right]$ & $\gamma_{i}$:~Instantaneous SNR             \\
  $\mathbb{E}[~.~]$:~Expectation operator & $\textrm{Pr}\{.\}$:~Probability of the given event               \\
  \hline
  \end{tabular}
\end{table}

\section{System Model and Assumptions}\label{model_Ch2}
\begin{figure} [t]
  \centering
  \includegraphics [width=5.25in] {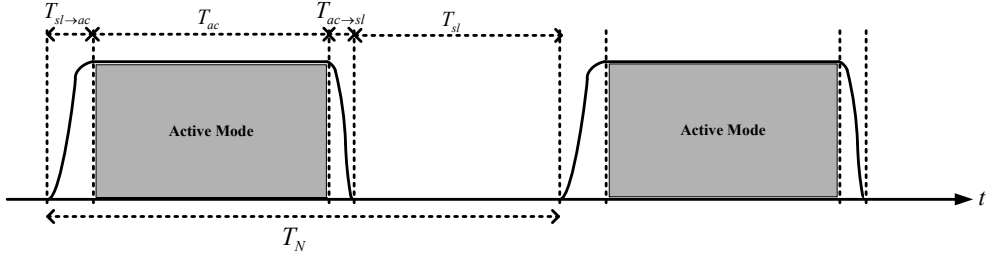}
  \caption{A practical duty-cycling process in a proactive WSN.}
  \label{fig: Time-Basis}
\end{figure}

In this work, we consider a proactive wireless sensor system, in
which a sensor node transmits an equal amount of data per time unit
to a designated sink node. The sensor and sink nodes synchronize
with one another and operate in a duty-cycling manner as depicted in
Fig \ref{fig: Time-Basis}. During \emph{active mode} period
$T_{ac}$, the sensed analog signal is first digitized by an
Analog-to-Digital Converter (ADC), and an $N$-bit binary message
sequence $\mathbb{M}_N \triangleq \lbrace{a_i \rbrace}_{i=1}^{N}$ is
generated, where $N$ is assumed to be fixed. The bit stream is
modulated using a pre-determined modulation scheme\footnote{Because
the main goal of this work is to find the distance-based green
modulations, and noting that the source/channel coding increase the
complexity and the power consumption, in particular codes with
iterative decoding process, the source/channel coding are not
considered.} and then transmitted to the sink node. Finally, the
sensor node returns to the \emph{sleep mode}, and all the circuits
are powered off for the sleep mode duration $T_{sl}$. We denote
$T_{tr}$ as the \emph{transient mode} duration consisting of the
switching time from sleep mode to active mode (i.e., $T_{sl
\rightarrow ac}$) plus the switching time from active mode to sleep
mode (i.e., $T_{ac \rightarrow sl}$), where $T_{ac \rightarrow sl}$
is short enough to be negligible. Under the above considerations,
the sensor/sink nodes have to process one entire $N$-bit message
$\mathbb{M}_N$ during $0 \leq T_{ac} \leq T_N-T_{sl}-T_{tr}$, where $T_N
\triangleq T_{tr}+T_{ac}+T_{sl}$ is assumed to be fixed for each
modulation, and $T_{tr} \approx T_{sl \rightarrow ac}$. Note that
$T_{ac}$ is an influential factor in choosing the energy-efficient
modulation, since it directly affects the total energy consumption
as we will show later.

We assume that both sensor and sink devices include a Direct Current
to Direct Current (DC-DC) converter to generate a desired supply
voltage from the embedded batteries. An DC-DC converter is specified
by its \emph{power transfer efficiency} denoted by $\chi_e <1$. In
addition, we assume a linear model for batteries with a small
discharge current, meaning that the stored energy will be completely
used or released. This model is reasonable for the proposed
duty-cycling process, as the batteries which are discharged during
active mode durations can recover their wasted capacities during
sleep mode periods. Since sensor nodes in a typical WSN are densely
deployed, the distance between nodes is normally short. Thus, the
total circuit power consumption, defined by $\mathcal{P}_c
\triangleq \mathcal{P}_{ct}+\mathcal{P}_{cr}$, is comparable to the
RF transmit power consumption denoted by $\mathcal{P}_t$, where
$\mathcal{P}_{ct}$ and $\mathcal{P}_{cr}$ represent the circuit
power consumptions for the sensor and sink nodes, respectively. In
considering the effect of the power transfer efficiency, the total
energy consumption in the active mode period, denoted by
$\mathcal{E}_{ac}$, is given by
\begin{equation}\label{active_energy1}
\mathcal{E}_{ac}=\dfrac{\mathcal{P}_c+\mathcal{P}_t}{\chi_e}T_{ac},
\end{equation}
where $T_{ac}$ is a function of $N$ and the channel bandwidth as we
will show in Section \ref{Analysis_Ch 3}. Also, it is shown in
\cite{Mingoo2007} that the power consumption during the sleep mode
duration $T_{sl}$ is much smaller than the power consumption in the
active mode (due to the low sleep mode leakage current) to be
negligible. As a result, the \emph{energy efficiency}, referred to
as the performance metric of the proposed WSN, can be measured by
the total energy consumption in each period $T_N$ corresponding to
$N$-bit message $\mathbb{M}_N$ as follows:
\begin{equation}\label{total_energy1}
\mathcal{E}_N \approx \frac{1}{\chi_e}\left[
(\mathcal{P}_c+\mathcal{P}_t)T_{ac}+\mathcal{P}_{tr}T_{tr}\right],
\end{equation}
where $\frac{\mathcal{P}_{tr}}{\chi_e}T_{tr}$ is the circuit energy
consumption during the transient mode period. We use
(\ref{total_energy1}) to investigate and compare the energy
efficiency of various modulation schemes in the subsequent sections.

\textbf{Channel Model:} It is shown that for short-range
transmissions including the wireless sensor networking, the root
mean square (rms) delay spread is in the range of ns
\cite{Karl_Book2005} (and ps for UWB applications
\cite{Tanchotikul2006}) which is small compared to the symbol
duration $T_s=16~\mu$s obtained from the bandwidth
$B=\frac{1}{T_s}=62.5$ KHz in the IEEE 802.15.4 standard \cite[p.
49]{IEEE_802_15_4_2006}. Thus, it is reasonable to expect a
flat-fading channel model for WSNs. Under the above considerations,
the channel model between the sensor and sink nodes is assumed to be
Rayleigh flat-fading with path-loss. This assumption is used in many
works in the literature (e.g., see \cite{Qu_ITSP0908} for WSNs). We
denote the fading channel coefficient corresponding to symbol $i$ as
$h_{i}$, where the amplitude $\big\vert h_{i} \big\vert$ is Rayleigh
distributed with the probability density function (pdf) $f_{\vert
h_{i} \vert}(r)=\frac{2r}{\Omega}e^{-\frac{r^2}{\Omega}},~r \geq 0$,
where $\Omega \triangleq \mathbb{E}\left[\vert h_{i} \vert^2\right]$
\cite{Proakis2001}. To model the path-loss of a link where the
transmitter and receiver are separated by distance $d$, let denote
$\mathcal{P}_t$ and $\mathcal{P}_r$ as the transmitted and the
received signal powers, respectively. For a $\eta^{th}$-power
path-loss channel, the channel gain factor is given by
\begin{equation}
\mathcal{L}_d \triangleq
\frac{\mathcal{P}_t}{\mathcal{P}_r}=M_ld^\eta
\mathcal{L}_1,~~~\textrm{with}~\eta_{min} \leq \eta \leq \eta_{max},
\end{equation}
where $M_l$ is the gain margin which accounts for the effects of
hardware process variations and $\mathcal{L}_1 \triangleq \frac{(4
\pi)^2}{\mathcal{G}_t \mathcal{G}_r \lambda^2}$ is the gain factor
at $d=1$ meter which is specified by the transmitter and receiver
antenna gains $\mathcal{G}_t$ and $\mathcal{G}_r$, and wavelength
$\lambda$ (e.g., \cite{Qu_ITSP0908}, \cite{Cui_GoldsmithITWC0905}).
As a result, when both fading and path-loss are considered, the
instantaneous channel coefficient becomes $G_{i} \triangleq
\frac{h_{i}}{\sqrt{\mathcal{L}_d}}$. Denoting $x_i(t)$ as the
transmitted signal with energy $\mathcal{E}_{t}$, the received
signal at the sink node is given by
$y_{i}(t)=G_{i}x_{i}(t)+z_{i}(t)$, where $z_{i}(t)$ is AWGN with
two-sided power spectral density given by $\frac{N_{0}}{2}$. Thus,
the instantaneous Signal-to-Noise Ratio (SNR) corresponding to an
arbitrary symbol $i$ can be computed as $\gamma_{i}=\frac{\vert
G_{i}\vert^2 \mathcal{E}_t}{N_0}$. Under the assumption of a
Rayleigh fading channel, $\gamma_{i}$ is chi-square distributed with
2 degrees of freedom, with pdf
$f_{\gamma}(\gamma_{i})=\frac{1}{\bar{\gamma}}\exp\left(
-\frac{\gamma_{i}}{\bar{\gamma}}\right)$, where $\bar{\gamma}
\triangleq \mathbb{E}[\vert
G_{i}\vert^2]\frac{\mathcal{E}_t}{N_0}=\frac{\Omega}{\mathcal{L}_d}\frac{\mathcal{E}_t}{N_0}$
denotes the average received SNR.

\section{Energy Efficiency Analysis of Sinusoidal Carrier-Based Modulations}\label{Analysis_Ch 3}
In this section, we analyze the energy and bandwidth efficiency of
three popular sinusoidal carrier-based modulations, namely MFSK,
MQAM and OQPSK, over a Rayleigh flat-fading channel with path-loss.
FSK is used in many low-complexity and energy-constrained wireless
systems and some IEEE standards (e.g., \cite{IEEE_P802_15}), whereas
MQAM is used in modem and digital video applications. Also, OQPSK is
used in the IEEE 802.15.4 standard which is the industry standard
for WSNs. In the sequel and for simplicity of the notation, we use
the superscripts `FS', `QA' and `OQ' for MFSK, MQAM and OQPSK,
respectively.

\textbf{M-ary FSK:} An M-ary FSK modulator with $M=2^b$ orthogonal
carriers benefits from the advantage of using the Direct Digital
Modulation (DDM) approach, meaning that it does not need the mixer
and the Digital to Analog Converter (DAC). This property makes MFSK
has a faster start-up time than the other modulation schemes. Let
denote $\Delta f=\frac{1}{\zeta T^{FS}_s}$ as the minimum carrier
separation with the symbol duration $T^{FS}_s$, where $\zeta=2$ for
coherent and $\zeta=1$ for non-coherent FSK \cite[p.
114]{Xiong_2006}. In this case, the channel bandwidth is obtained as
$B \approx M\times\Delta f $, where $B$ is assumed to be fixed for
all sinusoidal carrier-based modulations. Denoting $B^{FS}_{eff}$ as
the \emph{bandwidth efficiency} of MFSK defined as the ratio of data
rate $R^{FS}=\frac{b}{T^{FS}_s}$ (b/s) to the channel bandwidth, we
have
\begin{equation}\label{band_MFSK}
B^{FS}_{eff} \triangleq \frac{R^{FS}}{B} = \frac{\zeta \log_2
M}{M},~~~\textrm{b/s/Hz}.
\end{equation}
It can be seen that using a small constellation size $M$ avoid
losing more bandwidth efficiency in MFSK. To address the effect of
increasing $M$ on the energy efficiency, we first derive the
relationship between $M$ and the active mode duration $T^{FS}_{ac}$.
Since, we have $b$ bits during each symbol period $T^{FS}_{s}$, we
can write
\begin{equation}\label{active1}
T^{FS}_{ac}=\dfrac{N}{b}T^{FS}_{s}=\dfrac{MN}{\zeta B\log_2 M}.
\end{equation}
Recalling that $B$ and $N$ are fixed, an increase in $M$ results in
an increase in $T^{FS}_{ac}$. However, the maximum value of
$T^{FS}_{ac}$ is bounded by $T_N-T^{FS}_{tr}$ as illustrated in Fig.
\ref{fig: Time-Basis}. Thus, $M_{max}\triangleq 2^{b_{max}}$ in MFSK
is calculated by the following non-linear equation:
\begin{equation}\label{nonlinear}
\frac{M_{max}}{\log_2 M_{max}}=\frac{\zeta B}{N}(T_{N}-T^{FS}_{tr}).
\end{equation}

At the receiver side, the received MFSK signal can be detected
coherently to provide an optimum performance. However, the MFSK
coherent detection requires the receiver to obtain a precise
frequency and carrier phase reference for each of the transmitted
orthogonal carriers. For large $M$, this would increase the
complexity of the detector which makes a coherent MFSK receiver very
difficult to implement. Thus, most practical MFSK receivers use
non-coherent detectors\footnote{For the purpose of comparison, the
energy efficiency of a \emph{coherent} MFSK is fully analyzed in
Appendix I in \cite{JamshidTech2009}.}. To analyze the energy
efficiency of a NC-MFSK, we first derive $\mathcal{E}^{FS}_{t}$, the
transmit energy per symbol, in terms of a given average Symbol Error
Rate (SER) denoted by $P_{s}$. It is shown in \cite[Lemma
2]{TangITWC0407} that the average SER of a NC-MFSK is upper bounded
by
\begin{equation}
P_{s} = 1-\left(1-\frac{1}{2+\bar{\gamma}^{FS}}\right)^{M-1},
\end{equation}
where $\bar{\gamma}^{FS}
=\frac{\Omega}{\mathcal{L}_d}\frac{\mathcal{E}^{FS}_t}{N_0}$. As a
result, the transmit energy consumption per symbol is obtained from
the above $P_s$ as
\begin{equation}\label{energyFSK}
\mathcal{E}^{FS}_t \triangleq \mathcal{P}^{FS}_t T^{FS}_s =
\left[\left( 1-(1-P_s)^{\frac{1}{M-1}}\right)^{-1}-2
\right]\frac{\mathcal{L}_d N_0}{\Omega}.
\end{equation}
In considering the effect of the DC-DC converter and using
(\ref{active1}), the output energy consumption of transmitting
$N$-bit during $T^{FS}_{ac}$ is computed as
\begin{equation}\label{energy_trans1}
\dfrac{\mathcal{P}^{FS}_t}{\chi_e} T^{FS}_{ac} =
\dfrac{T^{FS}_{ac}}{T^{FS}_s}\dfrac{\mathcal{E}^{FS}_t}{\chi_e} =
\left[\left( 1-(1-P_s)^{\frac{1}{M-1}}\right)^{-1}-2
\right]\dfrac{\mathcal{L}_d N_0}{\chi_e \Omega} \dfrac{N}{\log_2 M}.
\end{equation}
On the other hand, the total circuit energy consumption of the
sensor/sink devices during $T^{FS}_{ac}$ is obtained from
$\frac{\mathcal{P}^{FS}_{ct}+\mathcal{P}^{FS}_{cr}}{\chi_e}T^{FS}_{ac}$.
For the sensor node with the MFSK modulator, we denote the power
consumption of frequency synthesizer, filters and power amplifier as
$\mathcal{P}^{FS}_{Sy}$, $\mathcal{P}^{FS}_{Filt}$ and
$\mathcal{P}^{FS}_{Amp}$, respectively. In this case,
\begin{equation}
\mathcal{P}^{FS}_{ct}=\mathcal{P}^{FS}_{Sy}+\mathcal{P}^{FS}_{Filt}+\mathcal{P}^{FS}_{Amp}.
\end{equation}
It is shown that the relationship between $\mathcal{P}^{FS}_{Amp}$
and the transmission power of an MFSK signal is
$\mathcal{P}^{FS}_{Amp}=\alpha^{FS} \mathcal{P}^{FS}_{t}$, where
$\alpha^{FS}$ is determined based on the type of the power
amplifier. For instance for a class B power amplifier,
$\alpha^{FS}=0.33$ \cite{Cui_GoldsmithITWC0905}. For the circuit
power consumption of the sink node, we use the fact that the optimum
NC-MFSK demodulator consists of a bank of $M$ matched filters, each
followed by an envelope detector \cite{ZiemerBook1985}. In addition,
we assume that the sink node uses a Low-Noise Amplifier (LNA) which
is generally placed at the front-end of a RF receiver circuit, an
Intermediate-Frequency Amplifier (IFA), and an ADC, regardless of
type of deployed modulation. Thus, denoting
$\mathcal{P}^{FS}_{LNA}$, $\mathcal{P}^{FS}_{Filr}$,
$\mathcal{P}^{FS}_{ED}$, $\mathcal{P}^{FS}_{IFA}$ and
$\mathcal{P}^{FS}_{ADC}$ as the power consumption of LNA, filters,
envelope detector, IF amplifier and ADC, respectively, the circuit
power consumption of the sink node is obtained as
\begin{equation}
\mathcal{P}^{FS}_{cr}=\mathcal{P}^{FS}_{LNA}+M\times(\mathcal{P}^{FS}_{Filr}+\mathcal{P}^{FS}_{ED})+\mathcal{P}^{FS}_{IFA}+\mathcal{P}^{FS}_{ADC}.
\end{equation}
Moreover, it is shown that the power consumption during the
transient mode period $T^{FS}_{tr}$ is governed by the frequency
synthesizer in both sensor/sink nodes \cite{Cui_GoldsmithITWC0905}.
Thus, the energy consumption during  $T^{FS}_{tr}$ is obtained as
$\frac{\mathcal{P}^{FS}_{tr}}{\chi_e}T^{FS}_{tr}=2
\frac{\mathcal{P}^{FS}_{Sy}}{\chi_e}T^{FS}_{tr}$
\cite{Karl_Book2005}. Substituting (\ref{active1}) and
(\ref{energy_trans1}) in (\ref{total_energy1}), the total energy
consumption of a NC-MFSK scheme for transmitting $N$-bit in each
period $T_N$ and for a given $P_s$ is obtained as
\begin{eqnarray}
\notag \mathcal{E}^{FS}_N &=&(1+\alpha^{FS}) \left(
\left[1-(1-P_s)^{\frac{1}{M-1}}\right]^{-1}-2 \right)
\dfrac{\mathcal{L}_d N_0}{\chi_e\Omega} \dfrac{N}{\log_2 M}+\\
\label{energy_totFSK}&&\dfrac{1}{\chi_e} \left[
(\mathcal{P}^{FS}_{c}-\mathcal{P}^{FS}_{Amp})\dfrac{MN}{B\log_{2}M}+2
\mathcal{P}^{FS}_{Sy}T^{FS}_{tr}\right],
\end{eqnarray}
with the fact that $\mathcal{L}_d=M_ld^\eta \mathcal{L}_1$. Thus,
the optimization goal is to determine the optimum constellation size
$M$, such that the objective function $\mathcal{E}^{FS}_N$ can be
minimized, i.e.,
\begin{eqnarray}
\notag \hat{M}=\textrm{arg}~\min_{M}~\mathcal{E}^{FS}_N~~~~~~~~~~~~~~~\\
\textrm{subject to} \left\{\begin{array}{l}
2\leq M \leq M_{max} \\
\eta_{min} \leq \eta \leq \eta_{max} \\
d >0,
\end{array} \right.
\end{eqnarray}
where $M_{max}$ derived from (\ref{nonlinear}). To solve this
optimization problem, we prove that (\ref{energy_totFSK}) is a
monotonically increasing function of $M$ for every value of $d$ and
$\eta$. It is seen that the second term in (\ref{energy_totFSK}) is
a monotonically increasing function of $M$. Also, from the first
term in (\ref{energy_totFSK}), we have
\begin{eqnarray}
\label{mono9}\left(\left[1-(1-P_s)^{\frac{1}{M-1}}\right]^{-1}-2\right) \dfrac{1}{\log_2 M}&=& \left(\left[1-e^{\frac{1}{M-1}\ln(1-P_s)}\right]^{-1}-2\right)\dfrac{1}{\log_2 M}\\
&\stackrel{(a)}{\approx}& \left(\left[1-e^{-\frac{P_s}{M-1}}\right]^{-1}-2\right)\dfrac{1}{\log_2 M}\\
\label{mono10}&\stackrel{(b)}{\approx}&\left(\dfrac{M-1}{P_s}-2\right)\dfrac{1}{\log_2 M},
\end{eqnarray}
where $(a)$ comes from the approximation $\ln(1-z)\approx -z,~~|z|
\ll 1$, and the fact that $P_s$ scales as $o(1)$. Also, $(b)$
follows from the approximation
$e^{-z}=\sum_{n=0}^{\infty}(-1)^{n}\frac{z^n}{n!}\approx
1-z,~~|z|\ll1$. It is concluded from (\ref{mono10}) that the first
term in (\ref{energy_totFSK}) is also a monotonically increasing
function of $M$. As a result, the minimum total energy consumption
$\mathcal{E}^{FS}_N$ is achieved at $\hat{M}=2$ for all values of
$d$ and $\eta$.

\textbf{M-ary QAM:} For $M$-ary QAM with the square constellation,
each $b=\log_{2}M$ bits of the message is mapped to the symbol
$S_i$, $i=0,1,...,M-1$, with the symbol duration $T^{QA}_s$.
Assuming the raised-cosine filter is used for the pulse shaping, the
channel bandwidth of MQAM is given by $B \approx
\frac{1}{2T^{QA}_s}$. Thus, using the data rate
$R^{QA}=\frac{b}{T^{QA}_s}$, the bandwidth efficiency of
 MQAM is obtained as $B^{QA}_{eff}\triangleq
\frac{R^{QA}}{B}=2\log_2M$ which is a logarithmically increasing
function of $M$. To address the impact of $M$ on the energy
efficiency, we derive the active mode duration $T^{QA}_{ac}$ in
terms of $M$ as follows:
\begin{equation}\label{active_QAM}
T^{QA}_{ac}=\dfrac{N}{b}T^{QA}_{s}=\dfrac{N}{2B\log_2M}.
\end{equation}
It is seen that an increase in $M$ results in a decrease in
$T^{QA}_{ac}$. Also compared to (\ref{active1}) for the NC-MFSK, it
is concluded that $T^{QA}_{ac}=\frac{1}{2M}T^{FS}_{ac}$.
Interestingly, it seems that the large constellation sizes $M$ would
result in the lower energy consumption due to the smaller values of
$T^{QA}_{ac}$. However, as we will show later, the total energy
consumption of a MQAM is not necessarily a monotonically decreasing
function of $M$. For this purpose, we obtain the transmit energy
consumption $\mathcal{P}^{QA}_t T^{QA}_{ac}$ with a similar argument
as for MFSK. It is shown in \cite[pp. 226]{Simon2005} and
\cite{JamshidTech2009} that the average SER of a coherent MQAM is
upper bounded by
\begin{equation}
P_{s} =
\dfrac{4(M-1)}{3\bar{\gamma}^{QA}+2(M-1)}\left(1-\dfrac{1}{\sqrt{M}}
\right),
\end{equation}
where $\bar{\gamma}^{QA}
=\frac{\Omega}{\mathcal{L}_d}\frac{\mathcal{E}^{QA}_t}{N_0}$ denotes
the average received SNR with the energy per symbol
$\mathcal{E}^{QA}_t$. As a result,
\begin{equation}
\mathcal{E}^{QA}_t \triangleq \mathcal{P}^{QA}_t T^{QA}_s =
\dfrac{2(M-1)}{3}\left[ 2 \left( 1-\dfrac{1}{\sqrt{M}}
\right)\dfrac{1}{P_s}-1 \right] \dfrac{\mathcal{L}_dN_0}{\Omega}.
\end{equation}
In considering the effect of the DC-DC converter, the output energy
consumption of transmitting $N$-bit during the active mode period is
computed as
\begin{equation}\label{energy_trans_MQAM}
\dfrac{\mathcal{P}^{QA}_t}{\chi_e} T^{QA}_{ac} =
\dfrac{T^{QA}_{ac}}{T^{QA}_s}\dfrac{\mathcal{E}^{QA}_t}{\chi_e} =
\dfrac{2(M-1)}{3}\left[ 2 \left( 1-\dfrac{1}{\sqrt{M}}
\right)\dfrac{1}{P_s}-1 \right] \dfrac{\mathcal{L}_dN_0}{\chi_e
\Omega} \dfrac{N}{\log_2 M},
\end{equation}
which is a monotonically increasing function of $M$ for every value
of $P_s$, $d$ and $\eta$. For the sensor node with the MQAM
modulator,
\begin{equation}
\mathcal{P}^{QA}_{ct}=\mathcal{P}^{QA}_{DAC}+\mathcal{P}^{QA}_{Sy}+\mathcal{P}^{QA}_{Mix}+\mathcal{P}^{QA}_{Filt}+\mathcal{P}^{QA}_{Amp},
\end{equation}
where $\mathcal{P}^{QA}_{DAC}$ and $\mathcal{P}^{QA}_{Mix}$ denote
the power consumption of DAC and mixer, respectively. It is shown
that $\mathcal{P}^{QA}_{Amp}=\alpha^{QA} \mathcal{P}^{QA}_{t}$,
where $\alpha^{QA}=\frac{\xi}{\vartheta}-1$ with
$\xi=3\frac{\sqrt{M}-1}{\sqrt{M}+1}$ and $\vartheta=0.35$
\cite{Cui_GoldsmithITWC0905}. In addition, the circuit power
consumption of the sink with the coherent MQAM is obtained as
\begin{equation}
\mathcal{P}^{QA}_{cr}=\mathcal{P}^{QA}_{LNA}+\mathcal{P}^{QA}_{Mix}+\mathcal{P}^{QA}_{Sy}+\mathcal{P}^{QA}_{Filr}+\mathcal{P}^{QA}_{IFA}+\mathcal{P}^{QA}_{ADC}.
\end{equation}
Also, with a similar argument as for MFSK, we assume that the
circuit power consumption during transient mode period $T^{QA}_{tr}$
is governed by the frequency synthesizer. As a result, the total
energy consumption of a coherent MQAM for transmitting $N$-bit in
each period $T_N$ is obtained as
\begin{eqnarray}
\notag \mathcal{E}^{QA}_N &=& (1+\alpha^{QA})
\dfrac{2(M-1)}{3}\left[ 2 \left( 1-\dfrac{1}{\sqrt{M}}
\right)\dfrac{1}{P_s}-1 \right]
\dfrac{\mathcal{L}_{d}N_0}{\chi_e\Omega} \dfrac{N}{\log_2 M}+\\
\label{energy_totMQAM}&&
\dfrac{1}{\chi_e}\left[(\mathcal{P}^{QA}_{c}-\mathcal{P}^{QA}_{Amp})\dfrac{N}{2B\log_{2}M}
+2 \mathcal{P}^{QA}_{Sy}T^{QA}_{tr}\right].
\end{eqnarray}
Although, there is no constraint on the maximum size $M$ for MQAM,
to make a fair comparison to the MFSK scheme, we use the same
$M_{max}$ as MFSK. Taking this into account, the optimization
problem is to determine the optimum $M \in [4,M_{max}]$ subject to
$d>0$ and $\eta_{min} \leq \eta \leq \eta_{max}$, such that
$\mathcal{E}^{QA}_N$ can be minimized.

It is seen that the first term in (\ref{energy_totMQAM}) is a
monotonically increasing function of $M$ for every value of $P_s$,
$d$ and $\eta$, while the second term is a monotonically decreasing
function of $M$ which is independent of $d$ and $\eta$. For the
above optimization and for a given $P_s$, we have two following
scenarios based on the distance $d$:

\textbf{Case 1:} For large values of $d$ where the first term in
(\ref{energy_totMQAM}) is dominant, the objective function
$\mathcal{E}^{QA}_N$ is a monotonically increasing function of $M$
and is minimized at $M=4$, equivalent to the 4-QAM scheme.

\textbf{Case 2:} Let assume that $d$ is small enough. One possible
case may happen is when the total energy consumption
$\mathcal{E}_{N}^{QA}$ for small sizes $M$ is governed by the second
term in (\ref{energy_totMQAM}). For this situation, either the
objective function behaves as a monotonically decreasing function of
$M$ for every value of $M$, or for a large constellation size $M$,
the first term would be dominant, meaning that
$\mathcal{E}_{N}^{QA}$ increases when $M$ grows. In the former
scenario, the optimum $M$ is achieved at $\hat{M}=M_{max}$; whereas
in the latter scenario, there exists a minimum value for
$\mathcal{E}_{N}^{QA}$, where the optimum $M$ for this point is
obtained by the intersection between the first and second terms,
i.e.,
\begin{equation}
(1+\alpha^{QA}) \dfrac{2(M-1)}{3}\left[ 2 \left(
1-\dfrac{1}{\sqrt{M}} \right)\dfrac{1}{P_s}-1 \right]
\dfrac{\mathcal{L}_{d}N_0}{\chi_e\Omega} \dfrac{N}{\log_2
M}=\dfrac{1}{\chi_e}\left[\dfrac{N(\mathcal{P}^{QA}_{c}-\mathcal{P}^{QA}_{Amp})}{2B\log_{2}M}
+2 \mathcal{P}^{QA}_{Sy}T^{QA}_{tr}\right]. \nonumber
\end{equation}
Since, $P_s$ scales as $o(1)$ and ignoring the term
$\mathcal{P}^{QA}_{Sy}T^{QA}_{tr}$ to simplify our analysis, the
optimum $M$ which minimizes (\ref{energy_totMQAM}) is obtained by
the following equation:
\begin{equation}\label{equ01}
M-1-\sqrt{M}+\dfrac{1}{\sqrt{M}} \approx
\dfrac{\phi(d,\eta)}{1+\alpha^{QA}},
\end{equation}
where $\phi(d,\eta) \triangleq
\frac{\mathcal{P}^{QA}_{c}-\mathcal{P}^{QA}_{Amp}}{2B}\frac{3P_s
\Omega}{4\mathcal{L}_{d}N_0}$ and the fact that $\alpha^{QA}$ is a
function of $M$.

\begin{table}
\label{table019} \caption{System Evaluation Parameters} \centering
  \begin{tabular}{|l|l|l|}
  \hline

   $\chi_e=0.8$          & $N_0=-180$ dBm                & $\mathcal{P}_{DAC}=7$ mw  \\

   $B=62.5$ KHz          & $\mathcal{P}_{ED}=3$ mw       & $\mathcal{P}_{ADC}=7$ mw  \\

   $M_l=40$ dB           & $\mathcal{P}_{Sy}=10$ mw      & $\mathcal{P}_{Mix}=7$ mw  \\

   $\mathcal{L}_1=30$ dB & $\mathcal{P}_{Filt}=2.5$ mw   &  $\mathcal{P}_{LNA}=9$ mw \\

   $\Omega=1$            & $\mathcal{P}_{Filr}=2.5$ mw   & $\mathcal{P}_{IFA}=3$ mw  \\

   \hline
  \end{tabular}
\end{table}

\begin{table}
\label{table041} \caption{Optimum $M$ in the MQAM optimization
problem for $P_s=10^{-3}$ and different values of $d$ and $2.5 \leq
\eta \leq6$} \centering
  \begin{tabular}{|c|c|c|c|c|c|}
  \hline
   $d$ (m) & $\eta=2.5$  & $\eta=3$  & $\eta=4$ & $\eta=5$ & $\eta=6$  \\
  \hline
    1      & 64        &   64      &  64      &   64     &  64   \\
   10      & 64        &   64      &  43      &   10     &  4    \\
   20      & 64        &   50      &  8       &   4      &  4    \\
   40      & 43        &   13      &  4       &   4      &  4    \\
   80      & 14        &   5       &  4       &   4      &  4    \\
   100     & 10        &   4       &  4       &   4      &  4    \\
   150     & 6         &   4       &  4       &   4      &  4    \\
   200     & 5         &   4       &  4       &   4      &  4    \\
  \hline
  \end{tabular}
\end{table}

\begin{figure}[t]
\centerline{\psfig{figure=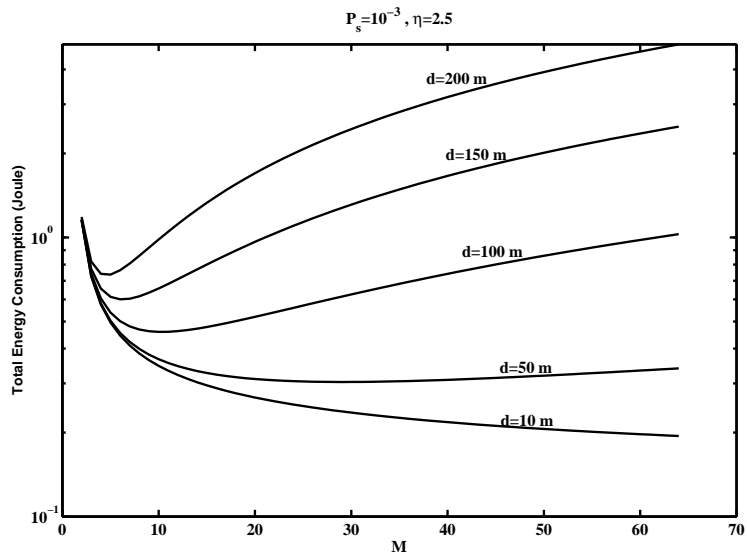,width=4.55in}}
\vspace{-7pt} \center{\hspace{16pt} \small{(a)}} \vspace{10pt}
\hspace{1pt}
\centerline{\psfig{figure=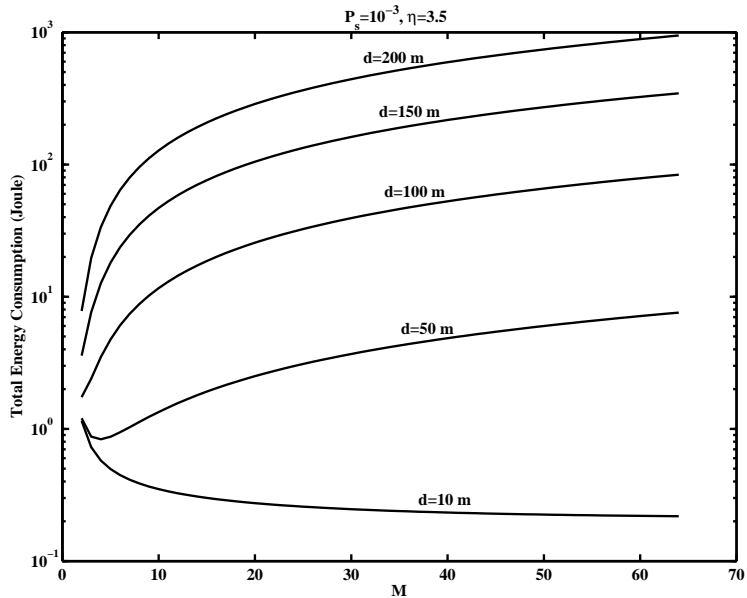,width=4.55in}}
\vspace{-35pt}
\center{\hspace{14pt} \small{(b)}} \\
\vspace{-7pt} \caption[a)  and b) .] { \small{Total energy
consumption $\mathcal{E}_{N}^{QA}$ vs. $M$ over a Rayleigh fading
channel with path-loss for $P_s=10^{-3}$, a) $\eta=2.5$ , and b)
$\eta=3.5$.}} \label{fig: Total_Energy}
\end{figure}

To gain more insight to the above optimization problem, we use a
specific numerical example with the simulation parameters summarized
in Table II\footnote{For more details in the simulation parameters,
we refer the reader to \cite{JamshidTech2009} and its references.}.
We assume $P_s=10^{-3}$, $4 \leq M \leq 64$ and $2.5 \leq \eta \leq
6$. Fig. \ref{fig: Total_Energy} illustrates the total energy
consumption of MQAM versus $M$ for different values of $d$. It is
seen that $\mathcal{E}_{N}^{QA}$ exhibits different trends depending
on the distance $d$ and the path-loss exponent $\eta$. For instance,
for large values of $d$ and $\eta$, $\mathcal{E}_{N}^{QA}$ is an
increasing function of $M$, where the optimum value of $M$ is
achieved at $\hat{M}=4$ as expected. This is because, in this case,
the first term in (\ref{energy_totMQAM}) corresponding to the RF
signal energy consumption dominates $\mathcal{E}_{N}^{QA}$. Table
III details the optimum values of $M$ which minimizes
$\mathcal{E}_{N}^{QA}$ for some values of $1 \leq d \leq 200$ m and
$2.5 \leq \eta \leq 6$. We use these results to compare the energy
efficiency of the optimized MQAM with the other schemes in the
subsequent sections.

\textbf{Offset-QPSK:} For performance comparison, we choose the
conventional OQPSK modulation which is used as a reference in the
IEEE 802.15.4/ZigBee protocols. We also follow the same differential
OQPSK structure mentioned in \cite[p. 50]{IEEE_802_15_4_2006} to
eliminate the need for a coherent phase reference at the sink node.
For this configuration, the channel bandwidth and the data rate are
determined by $B \approx \frac{1}{T^{OQ}_s}$ and
$R^{OQ}=\frac{2}{T^{OQ}_s}$, respectively. As a result, the
bandwidth efficiency of OQPSK is obtained as $B^{OQ}_{eff}
\triangleq \frac{R^{OQ}}{B}=2$ (b/s/Hz). Since we have 2 bits in
each symbol period $T^{OQ}_{s}$, it is concluded that
\begin{equation}
T^{OQ}_{ac}=\frac{N}{2}T^{OQ}_{s}=\frac{N}{2B}.
\end{equation}
Compared to (\ref{active1}) and (\ref{active_QAM}), we have
$T^{OQ}_{ac}=(\log_2 M) T^{QA}_{ac}$ with $M \geq 4$, while for the
optimized MFSK, $T^{OQ}_{ac}=\frac{1}{4}T^{FS}_{ac}$. More
precisely, it is revealed that $T^{QA}_{ac} < T^{OQ}_{ac} <
T^{FS}_{ac}$. To determine the transmit energy consumption of the
differential OQPSK scheme, we derive $\mathcal{E}^{OQ}_t$ in terms
of the average SER. It is shown in \cite{SimonITC0603} and
\cite{JamshidTech2009} that the average SER of the differential
OQPSK is upper bounded by
\begin{equation}
P_{s}=\sqrt{\dfrac{1+\sqrt{2}}{2}}\dfrac{4}{(2-\sqrt{2})\bar{\gamma}^{OQ}+4},
\end{equation}
where
$\bar{\gamma}^{OQ}=\frac{\Omega}{\mathcal{L}_d}\frac{\mathcal{E}^{OQ}_t}{N_0}$.
With a similar argument as for MFSK and MQAM, the energy consumption
of transmitting $N$-bit during $T^{OQ}_{ac}$ is computed as
\begin{equation}\label{energy_trans_OQPSK}
\dfrac{\mathcal{P}^{OQ}_t}{\chi_e} T^{OQ}_{ac} =
\dfrac{T^{OQ}_{ac}}{T^{OQ}_s}\dfrac{\mathcal{E}^{OQ}_t}{\chi_e} =
\left[\dfrac{1}{2-\sqrt{2}} \left(
\dfrac{4}{P_s}\sqrt{\dfrac{1+\sqrt{2}}{2}}-4 \right)
\right]\dfrac{\mathcal{L}_dN_0}{\chi_e \Omega}\dfrac{N}{2}.
\end{equation}

In addition, for the sensor node with the OQPSK modulator,
$\mathcal{P}^{OQ}_{ct} \approx
\mathcal{P}^{OQ}_{DAC}+\mathcal{P}^{OQ}_{Sy}+\mathcal{P}^{OQ}_{Mix}+\mathcal{P}^{OQ}_{Filt}+\mathcal{P}^{OQ}_{Amp}$,
where we assume that the power consumption of the differential
encoder is negligible, and $\mathcal{P}^{OQ}_{Amp}=\alpha^{OQ}
\mathcal{P}^{OQ}_{t}$, with $\alpha^{OQ}=0.33$. In addition, the
circuit power consumption of the sink with the differential
detection OQPSK is obtained as
$\mathcal{P}^{OQ}_{cr}=\mathcal{P}^{OQ}_{LNA}+\mathcal{P}^{OQ}_{Mix}+\mathcal{P}^{OQ}_{Sy}+\mathcal{P}^{OQ}_{Filr}+\mathcal{P}^{OQ}_{IF}+\mathcal{P}^{OQ}_{ADC}$.
As a result, the total energy consumption of a differential OQPSK
system for transmitting $N$-bit in each period $T_N$ is obtained as
\begin{equation}\label{energy_totOQPSK}
\mathcal{E}^{OQ}_N = (1+\alpha^{OQ}) \left[\dfrac{1}{2-\sqrt{2}}
\left( \dfrac{4}{P_s}\sqrt{\dfrac{1+\sqrt{2}}{2}}-4 \right)
\right]\dfrac{\mathcal{L}_dN_0}{\chi_e \Omega}\dfrac{N}{2}
+\dfrac{1}{\chi_e}\left[(\mathcal{P}^{OQ}_{c}-\mathcal{P}^{OQ}_{Amp})\dfrac{N}{2B}+2
\mathcal{P}^{OQ}_{Sy}T^{OQ}_{tr}\right].
\end{equation}

\section{Numerical Results}\label{simulation_Ch5}
In this section, we present some numerical evaluations using
realistic parameters from the IEEE 802.15.4 standard and
state-of-the art technology to confirm the energy efficiency
analysis discussed in Section \ref{Analysis_Ch 3}. We assume that
all the modulations operate in the carrier frequency $f_0=$2.4 GHz
Industrial Scientist and Medical (ISM) unlicensed band utilized in
the IEEE 802.15.4 standard \cite{IEEE_802_15_4_2006}. According to
the FCC 15.247 RSS-210 standard for United States/Canada, the
maximum allowed antenna gain is 6 dBi \cite{FreeScale2007}. In this
work, we assume that $\mathcal{G}_t=\mathcal{G}_r=5$ dBi. Thus for
the $f_0=$2.4 GHz, $\mathcal{L}_1~ \textrm{(dB)}\triangleq
10\log_{10}\left(\frac{(4 \pi)^2}{\mathcal{G}_t \mathcal{G}_r
\lambda^2}\right) \approx 30~ \textrm{dB}$, where $\lambda
\triangleq \frac{3\times 10^8}{f_0}=0.125$ m. We assume that in each
period $T_N$, the data frame $N=1024$ bytes (or equivalently
$N=8192$ bits) is generated for transmission for all the
modulations, where $T_N$ is assumed to be 1.4 s. The channel
bandwidth is set to the $B=62.5$ KHz, according to the IEEE 802.15.4
standard \cite[p. 49]{IEEE_802_15_4_2006}. In addition, we assume
that the path-loss exponent is in the range of $2.5$ to $6$
\footnote{ $\eta=2$ is regarded as a reference state for the
propagation in free space and is unattainable in practice. Also,
$\eta=4$ is for relatively lossy environments, and for indoor
environments, the path-loss exponent can reach values in the range
of 4 to 6.}. We use the system parameters summarized in Table II for
simulations. It is concluded from $ \frac{M_{max}}{\log_2
M_{max}}=\frac{\zeta B}{N}(T_{N}-T^{FS}_{tr})$ that $M_{max} \approx
64$ (or equivalently $b_{max} \approx 6$) for NC-MFSK. Since, there
is no constraint on the maximum $M$ in MQAM, we choose $4 \leq M
\leq 64$ for MQAM to be consistent with MFSK.

\begin{figure}[bhpt]
\centerline{\psfig{figure=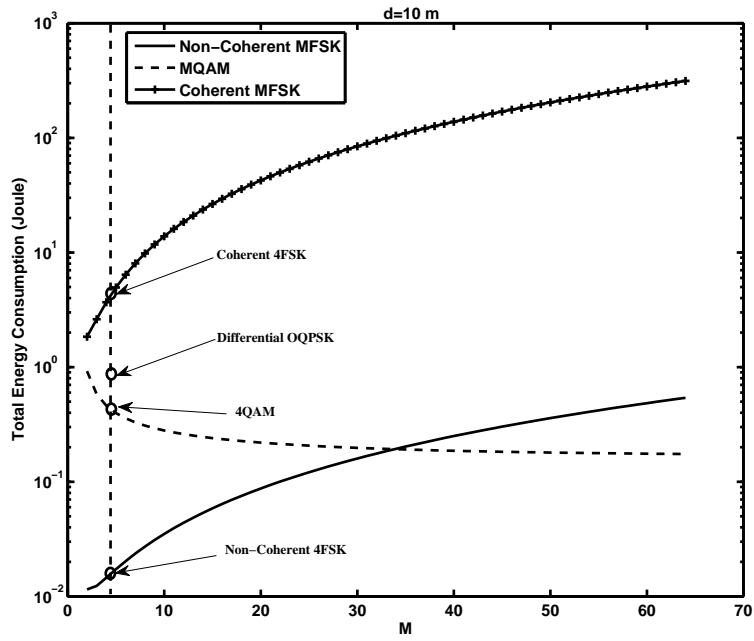,width=4.56in}}
\vspace{-7pt} \center{\hspace{16pt} \small{(a)}} \vspace{10pt}
\hspace{1pt}
\centerline{\psfig{figure=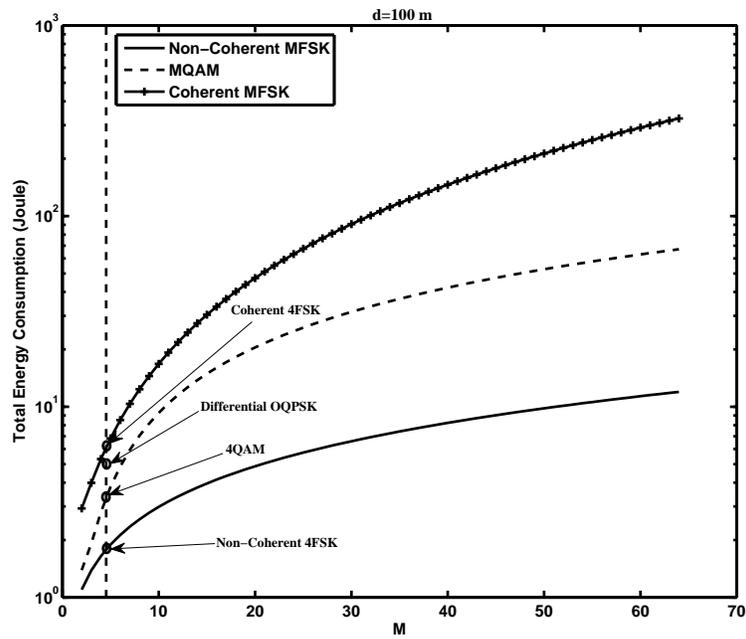,width=4.55in}}
\vspace{-35pt}
\center{\hspace{14pt} \small{(b)}} \\
\vspace{-7pt} \caption[a)  and b) .] { \small{Total energy
consumption of transmitting $N$-bit vs. $M$ for MFSK, MQAM and
differential OQPSK over a Rayleigh fading channel with path-loss and
$P_s=10^{-3}$, a) $d=10$ m, and b) $d=100$ m.}} \label{fig:
Total_Energy_dm}
\end{figure}

Fig. \ref{fig: Total_Energy_dm} compares the energy efficiency of
the modulation schemes investigated in Section \ref{Analysis_Ch 3}
versus $M$ for $P_s=10^{-3}$, $\eta=3.5$ and different values of
$d$. It is revealed from Fig. \ref{fig: Total_Energy_dm}-a that for
$M < 35$, NC-MFSK is more energy-efficient than MQAM, differential
OQPSK and coherent MFSK for $d=10$ m and $\eta=3.5$, while when $M$
grows, 64-QAM outperforms the other schemes for $d=10$ m. The latter
result is well supported by the Case 2 in the MQAM optimization
discussed in Section \ref{Analysis_Ch 3}. However, NC-MFSK for a
small size $M$ benefits from the advantage of less complexity and
cost in implementation than 64-QAM. Furthermore, as shown in Fig.
\ref{fig: Total_Energy_dm}-b, the total energy consumption of both
MFSK and MQAM for large $d$ increase logarithmically with $M$ which
verify the optimization solutions for the NC-MFSK and Case 1 for the
MQAM in Section \ref{Analysis_Ch 3}. Also, it is seen that NC-MFSK
exhibits the energy efficiency better than the other schemes when
$d$ increases.

The optimized modulations for different transmission distance $d$
and $2.5 \leq \eta \leq 6$ are listed in Table IV. For these
results, we use the optimized MQAM detailed in Table III and the
fact that for NC-MFSK, $\hat{M}=2$ is the optimum value which
minimizes $\mathcal{E}^{FS}_N$ for every $d$ and $\eta$. From Table
IV, it is found that although 64-QAM outperforms NC-BFSK for very
short range WSNs, it should be noted that using MQAM with a large
constellation size $M$ increases the complexity of the system. In
particular, when we know that MQAM utilizes the coherent detection
at the sink node. In other words, there exists a trade-off between
the complexity and the energy efficiency in using MQAM for small
values of $d$.

\begin{table}
\label{table0401} \caption{Energy-efficient modulation for
$P_s=10^{-3}$ and different values of $d$ and $\eta$} \centering
  \begin{tabular}{|c|c|c|c|c|c|}
  \hline
   $d$ (m) & $\eta=2.5$     & $\eta=3$        & $\eta=4$       & $\eta=5$       & $\eta=6$    \\
  \hline
    1      & 64QAM          &   64QAM         &  64QAM         &   64QAM        &  64QAM      \\
   10      & 64QAM          &   64QAM         &  NC-BFSK       &   NC-BFSK      &  NC-BFSK    \\
   20      & 64QAM          &   NC-BFSK       &  NC-BFSK       &   NC-BFSK      &  NC-BFSK    \\
   40      & NC-BFSK        &   NC-BFSK       &  NC-BFSK       &   NC-BFSK      &  NC-BFSK    \\
   80      & NC-BFSK        &   NC-BFSK       &  NC-BFSK       &   NC-BFSK      &  NC-BFSK    \\
   100     & NC-BFSK        &   NC-BFSK       &  NC-BFSK       &   NC-BFSK      &  NC-BFSK    \\
   150     & NC-BFSK        &   NC-BFSK       &  NC-BFSK       &   NC-BFSK      &  NC-BFSK    \\
   200     & NC-BFSK        &   NC-BFSK       &  NC-BFSK       &   NC-BFSK      &  NC-BFSK    \\
  \hline
  \end{tabular}
\end{table}

Up to know, we have investigated the energy efficiency of the
sinusoidal carrier-based modulations under the assumption of a
Rayleigh fading channel with path-loss. It is also of interest to
evaluate the energy efficiency of the aforementioned modulation
schemes operating over the more general Rician model which includes
the LOS path. For this purpose, let assume that the instantaneous
channel coefficient correspond to symbol $i$ is
$G_i=\frac{h_i}{\mathcal{L}_d}$, where $h_i$ is assumed to be Rician
distributed with pdf $f_{h_i}(r)=\frac{r}{\sigma^2}e^{-\frac{r^2 +
A^2}{2 \sigma^2} }I_{0} \left( \frac{rA}{\sigma^2} \right),~r \geq
0$, where $A$ denotes the peak amplitude of the dominant signal,
$2\sigma^2 \triangleq \Omega$ is  the average power of non-LOS
multipath components \cite[p. 78]{Goldsmith_Book2005}. For this
model, $\mathbb{E} [\vert h_i \vert ^2]=A^2 +2 \sigma^2 =2 \sigma^2
(1+K)$, where $K(\textrm{dB}) \triangleq 10 \log \frac{A^2}{2
\sigma^2}$ is the Rician factor. The value of $K$ is a measure of
the severity of the fading. For instance, $K(\textrm{dB}) \to
-\infty$ implies Rayleigh fading and $K(\textrm{dB}) \to \infty$
represents AWGN channel. Table V summarized the energy efficiency
results of the previous modulations over a Rician fading channel
with path-loss for $P_s=10^{-3}$ and $\eta=3.5$. It is seen from
Table V that NC-MFSK with a small size $M$ has less total energy
consumption than the other schemes in Rician fading channel with
path-loss.

\begin{table}
\label{table022} \caption{Total Energy Consumption (in Joule) of
NC-MFSK, MQAM and OQPSK over a Rician Fading Channel with Path-Loss
for $P_s=10^{-3}$ and $\eta=3.5$} \centering
  \begin{tabular}{|c|cccc|ccc|ccc|}
   \hline

   &          &        & $K=1$ dB &       &            & $K=10$ dB &       &        & $K=15$ dB &       \\
   \hline
   &  $M$     & OQPSK  & NC-MFSK  & MQAM  &   OQPSK    & NC-MFSK   & MQAM  & OQPSK  & NC-MFSK   & MQAM  \\
   \hline
   & 4        & 1.1241 & 0.0173   & 0.5621&    1.1241  & 0.0171    &0.5620 &  1.1241& 0.0171    & 0.5620\\

   d=10 m & 16         &        & 0.0769   & 0.2819&            & 0.0765    &0.2810 &        & 0.0765    & 0.2810\\

   & 64       &        & 0.6558   & 0.1924&            & 0.6545    &0.1874 &        & 0.6545    & 0.1874\\
   \hline
   \hline
   & 4          & 1.2236 & 0.5835   & 0.8873 &   1.1445   & 0.0194    &0.5652 & 1.1310 & 0.0175    & 0.5627\\

   d=100 m & 16         &        & 1.4920   & 3.2049 &            & 0.0785    &0.2989 &        & 0.0767    & 0.2843\\

   & 64         &        & 4.6199   & 16.1010&            & 0.6570    &0.2615 &        & 0.6547    & 0.2002\\
   \hline
  \end{tabular}
\end{table}

\begin{figure}[t]
\centerline{\psfig{figure=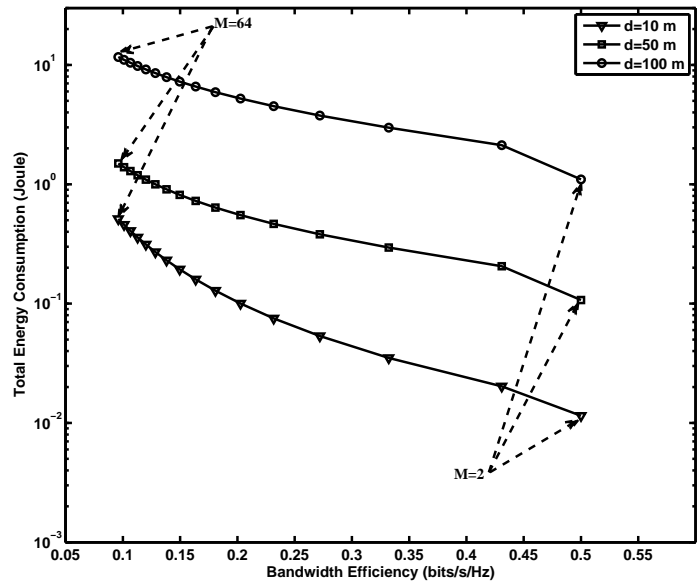,width=4.25in}}
\caption{Total energy consumption of transmitting $N$-bit versus
 bandwidth efficiency for NC-MFSK, and for $P_s=10^{-3}$ and different values of $d$ and $M$.} \label{fig: Energy_vs_Bandwidth}
\end{figure}

The above results make NC-MFSK with a small $M$ attractive for using
in WSNs, since this modulation already has the advantage of less
complexity and cost in implementation than MQAM, differential OQPSK
and coherent MFSK, and has less total energy consumption. In
addition, since for energy-constrained WSNs, data rates are usually
low, using M-ary NC-FSK schemes with a small $M$ are desirable. The
sacrifice, however, is the bandwidth efficiency of NC-MFSK (when $M$
increases) which is a critical factor in band-limited WSNs. Since
most of WSN applications operate in unlicensed bands where large
bandwidth is available, NC-MFSK can surpass the spectrum constraint
in WSNs. To have more insight into the above discussions for the
NC-MFSK, we plot the total energy consumption of NC-MFSK as a
function of $B_{eff}^{FS}$ for different values of $M$ and $d$ in
Fig. \ref{fig: Energy_vs_Bandwidth}. In all cases, we observe that
the minimum $\mathcal{E}_{N}^{FS}$ is achieved at low values of
distance $d$ and for $M=2$, which corresponds to the maximum
bandwidth efficiency $B_{eff}^{FS}=0.5$.

The analysis and numerical evaluations so far implicitly focused on
the sinusoidal carrier-based modulations with the bandwidth $B=62.5$
KHz. To complete our analysis, it is of interest to compare the
energy efficiency of the optimized NC-MFSK with the On-Off Keying
(OOK), known as the simplest UWB modulation scheme. For this
purpose, we first derive the total energy consumption of OOK with a
similar manner as for sinusoidal carrier-based modulations. Then, we
evaluate the energy efficiency of OOK in terms of distance $d$. For
simplicity of the notation, we use the superscript `OK' for OOK
modulation scheme.

\section{Energy Consumption Analysis of OOK}\label{Analysis_Ch 4}
For OOK, the number of bits per symbol is defined as $b=\log_2 M=1$.
An OOK transmitted signal corresponding to the symbol $a_i \in
\mathbb{M}_N$ is given by $x^{OK}_i(t)=\sqrt{\mathcal{E}^{OK}_t}a_i
p(t-iT^{OK}_s)$, where $p(t)$ is an ultra-short pulse of width $T_p$
with unit energy, $\mathcal{E}^{OK}_t$ is the transmit energy
consumption per symbol, and $T_{s}^{OK}$ is the OOK symbol duration.
The ratio $\frac{T_{p}}{T^{OK}_s}$ is defined as the
\emph{duty-cycle factor} of an OOK signal, which is the fractional
on-time of the OOK ``1'' pulse. The channel bandwidth and the data
rate of an OOK are determined as $B \approx \frac{1}{T_{p}}$ and
$R^{OK}=\frac{1}{T_{s}^{OK}}$, respectively. As a result, the
bandwidth efficiency of an OOK is obtained as $B^{OK}_{eff}
\triangleq \frac{R^{OK}}{B}=\frac{T_p}{T^{OK}_s} \leq 1$ (b/s/Hz)
which controls by the duty-cycle factor. Note that during the
transmission of the OOK ``0'' pulse, the filter and the power
amplifier of the OOK modulator are powered off. During this time,
however, the receiver is turned on to detect zero pulses. For this
reason, we still use the same definition for active mode period
$T^{OK}_{ac}$ as used for the sinusoidal carrier-based modulations
as follows:
\begin{equation}
T^{OK}_{ac}=\frac{N}{b}T^{OK}_{s}=N T^{OK}_{s}.
\end{equation}
Depend upon the duty-cycle factor, $T_{ac}^{OK}$ can be expressed in
terms of bandwidth $B$. For instance, for an OOK with the duty-cycle
factor $\frac{T_{p}}{T^{OK}_s}=\frac{1}{2}$, we have $T^{OK}_{ac}=2N
T_{p}=\frac{2N}{B}$, and $B^{OK}_{eff}=\frac{1}{2}$. Compared to (\ref{band_MFSK})
for the NC-MFSK, it is concluded that $B^{OK}_{eff}$ with the duty-cycle
factor $\frac{T_{p}}{T^{OK}_s}=\frac{1}{2}$ is the same as that of the
optimized MFSK (i.e., $\hat{M}=2$) in Section III.
It is shown in
\cite[pp. 490-504]{Couch_Book2001} that the average SER of an OOK
with non-coherent detection is upper bounded by
\begin{equation}
P_{s}=\dfrac{1}{\bar{\gamma}^{OK}+2},
\end{equation}
where
$\bar{\gamma}^{OK}=\frac{\Omega}{\mathcal{L}_d}\frac{\mathcal{E}^{OK}_t}{N_0}$
denotes the average received SNR. Thus, the transmit energy
consumption per symbol is obtained as $\mathcal{E}^{OK}_t \triangleq
\mathcal{P}^{OK}_t T_p = \left(\frac{1}{P_s}-2
\right)\frac{\mathcal{L}_dN_0}{\Omega}$ which corresponds to
transmitting ``1'' pulse. An interesting point is that
$\mathcal{E}^{OK}_t$ scales the same as $\mathcal{E}^{FS}_t$
obtained in (\ref{energyFSK}) for NC-BFSK. This can be easily proved
by using the approximation method in (\ref{mono9})-(\ref{mono10}).
It should be noted that the energy consumption of transmitting
$N$-bit during the active mode period, denoted by
$\frac{\mathcal{P}^{OK}_t}{\chi_e} T^{OK}_{ac}$, is equivalent to
the energy consumption of transmitting $L$-bit ``1'' in
$\mathbb{M}_N$, where $L$ is a binomial random variable with
parameters $(N,q)$. Assuming uncorrelated and equally likely binary
data $a_i$, we have $q=\frac{1}{2}$. Hence,
$\frac{\mathcal{P}^{OK}_t}{\chi_e} T^{OK}_{ac} =\frac{L}{\chi_e}
\mathcal{E}^{OK}_t = L \left(\frac{1}{P_s}-2
\right)\frac{\mathcal{L}_dN_0}{\chi_e \Omega}$, where $L$ has the
probability mass function $\textrm{Pr}\{ L=\ell
\}=\binom{N}{\ell}\left(\frac{1}{2}\right)^N$ with
$\mathbb{E}[L]=\frac{N}{2}$.

\begin{figure}[t]
\centerline{\psfig{figure=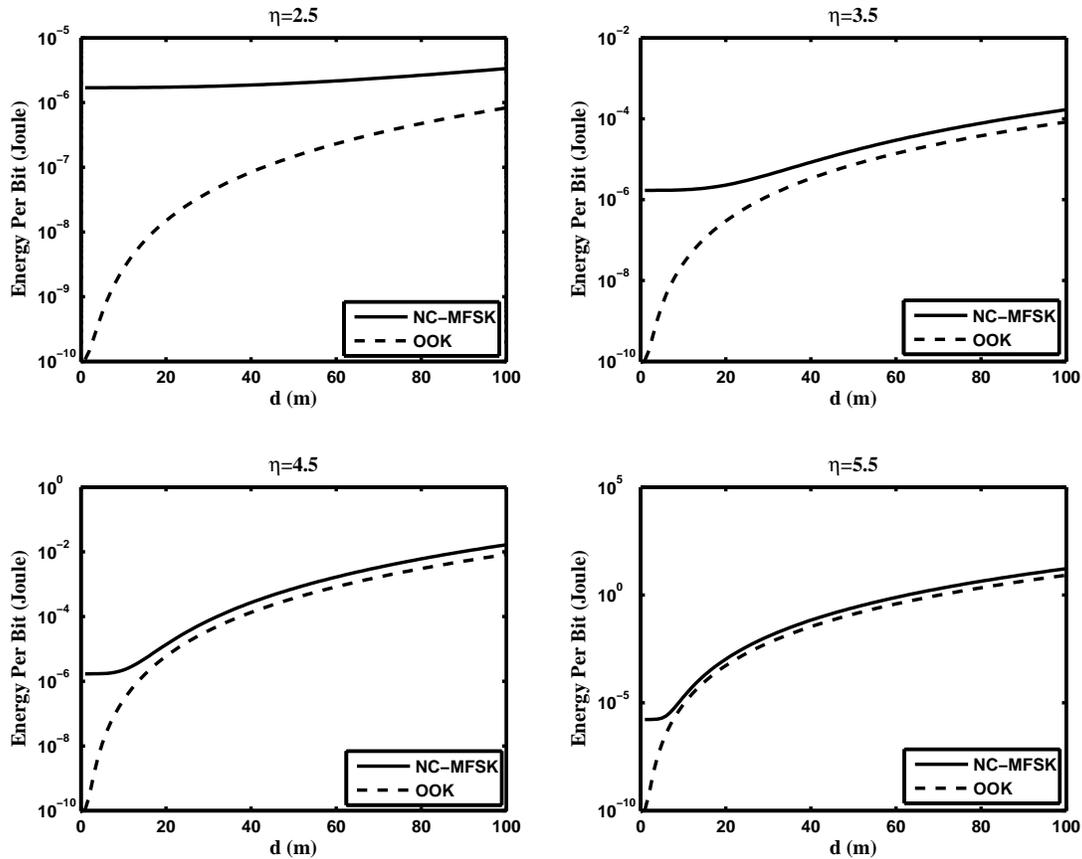,width=6.65in}}
\caption{Total energy consumption per information bit versus
 $d$ for OOK and the optimized NC-MFSK, and for $P_s=10^{-3}$.} \label{fig: Total_Energy_UWB_vs_d}
\end{figure}

We denote the power consumption of pulse generator, power amplifier
and filter as $\mathcal{P}^{OK}_{PG}$, $\mathcal{P}^{OK}_{Amp}$ and
$\mathcal{P}^{OK}_{Filt}$, respectively. Hence, the circuit energy
consumption of the sensor node during $T^{OK}_{ac}$ is represented
as a function of the random variable $L$ as
$\mathcal{P}^{OK}_{ct}T^{OK}_{ac} =
\mathcal{P}^{OK}_{PG}T^{OK}_{ac}+ L T_p
\left(\mathcal{P}^{OK}_{Filt}+\mathcal{P}^{OK}_{Amp}\right)$, where
the factor $L T_p $ comes from the fact that the filter and the
power amplifier are active only during the transmission of $L$-bit
``1''. We assume that $\mathcal{P}^{OK}_{Amp}=\alpha^{OK}
\mathcal{P}^{OK}_{t}$ with $\alpha^{OK}=0.33$. In addition, the
circuit energy consumption of the sink node with a non-coherent
detection during $T^{OK}_{ac}$ is obtained as
$\mathcal{P}^{OK}_{cr}T^{OK}_{ac}=\left(\mathcal{P}^{OK}_{LNA}+\mathcal{P}^{OK}_{ED}+\mathcal{P}^{OK}_{Filr}+\mathcal{P}^{OK}_{Int}+\mathcal{P}^{OK}_{ADC}\right)
T^{OK}_{ac}$, where $\mathcal{P}^{OK}_{Int}$ is the power
consumption of the integrator. With a similar argument as for the
sinusoidal carrier-based modulations, we assume that the circuit
power consumption during $T^{OK}_{tr}$ is governed by the pulse
generator. As a result, the total energy consumption of a
non-coherent OOK for transmitting $N$-bit is obtained as a function
of the random variable $L$ as follows:
\begin{equation}\label{energy_totOOK1}
\mathcal{E}^{OK}_N (L)=  (1+\alpha^{OK})L \left(\dfrac{1}{P_s}-2
\right)\dfrac{\mathcal{L}_dN_0}{\chi_e\Omega}
+\frac{1}{\chi_e}\left[(\mathcal{P}^{OK}_{cr}+\mathcal{P}^{OK}_{PG})\dfrac{2N}{B}+\dfrac{L}{B}\mathcal{P}^{OK}_{Filt}+2
\mathcal{P}^{OK}_{PG}T^{OK}_{tr}\right],
\end{equation}
where we use $T^{OK}_{ac}=\frac{2N}{B}$. Since
$\mathbb{E}[L]=\frac{N}{2}$, the average $\mathcal{E}^{OK}_N (L)$ is
computed as
\begin{eqnarray}
\notag\mathcal{E}^{OK}_N & \triangleq &
\mathbb{E}\left[\mathcal{E}^{OK}_N (L)\right]= (1+\alpha^{OK})
\left(\dfrac{1}{P_s}-2
\right)\dfrac{\mathcal{L}_dN_0}{\chi_e\Omega}\dfrac{N}{2}+\\
\label{energy_totOOK01}&&\frac{1}{\chi_e}\left[(\mathcal{P}^{OK}_{cr}+\mathcal{P}^{OK}_{PG})\dfrac{2N}{B}+\dfrac{N}{2B}\mathcal{P}^{OK}_{Filt}+2
\mathcal{P}^{OK}_{PG}T^{OK}_{tr}\right].
\end{eqnarray}
\begin{table}
\label{table01} \caption{OOK System Evaluation Parameters}
\centering
  \begin{tabular}{|l|l|l|}
  \hline
   $N=20000$   &   $N_0=-180$ dB    &  $\mathcal{P}_{PG}=675$ $\mu$w \\

   $B=500$ MHz &   $T_N=100$ msec   &  $\mathcal{P}_{LNA}=3.1$ mw    \\

   $M_l=40$ dB &   $T_{tr}=2$  nsec &  $\mathcal{P}_{ED}=3$ mw       \\

   $\mathcal{L}_1=30$ dB & $\mathcal{P}_{Filt}=2.5$ mw  & $\mathcal{P}_{ADC}=7$ mw \\

   $\chi_e=0.8$   &   $\mathcal{P}_{Filr}=2.5$ mw   & $\mathcal{P}_{Int}=3$ mw  \\
   \hline
  \end{tabular}
\end{table}
It should be noted that the OOK scheme uses the channel bandwidth
much wider than that of the sinusoidal carrier-based modulations.
Thus, to make a fair comparison with the optimized NC-MFSK, it is
reasonable to use the total energy consumption per information bit
defined as $\mathcal{E}_b \triangleq \frac{\mathcal{E}_N}{N}$,
instead of using $\mathcal{E}_N$. In our comparison, we use the
simulation parameters shown in Table VI \cite{JamshidTech2009}. Fig.
\ref{fig: Total_Energy_UWB_vs_d} compares the total energy
consumption per information bit of the OOK with that of the
optimized NC-MFSK versus communication range $d$ for $P_s=10^{-3}$
and different values of $\eta$. We observe that a significant energy
saving is achieved using OOK as compared to the optimized NC-MFSK,
when $d$ and $\eta$ decrease. While, for the indoor environments
where $\eta$ is large, the performance difference between OOK and
the optimized NC-MFSK vanishes as $d$ increases. This is because,
$i)$ the transmission energy consumption in the active mode period
is dominant when $d$ increases, $ii)$ $\mathcal{E}^{OK}_t$ scales
the same as $\mathcal{E}^{FS}_t$ obtained in (\ref{energyFSK}) using
(\ref{mono10}). Since, UWB modulation schemes use the channel bandwidth
much wider than that of the sinusoidal carrier-based modulations, the
optimized NC-MFSK is desirable in use for the band-limited and sparse WSNs
where the path-loss exponent is large.

\section{Conclusion}\label{conclusion_Ch6}
In this paper, we have analyzed the energy efficiency of some
popular modulation schemes to find the distance-based green
modulations in a WSN over Rayleigh and Rician flat-fading channels
with path-loss. It was demonstrated that among various sinusoidal
carrier-based modulations, the optimized NC-MFSK is the most
energy-efficient scheme in sparse WSNs for each value of the
path-loss exponent, where the optimization is performed over the
modulation parameters. In addition, NC-MFSK with a small $M$ is
attractive for using in WSNs, since this modulation already has the
advantage of less complexity and cost in implementation than MQAM,
differential OQPSK and coherent MFSK, and has less total energy
consumption. Furthermore, MFSK has a faster start-up time than other
schemes. Moreover, since for energy-constrained WSNs, data rates are
usually low, using M-ary NC-FSK schemes with a small $M$ are
desirable. The sacrifice, however, is the bandwidth efficiency of
NC-MFSK when $M$ increases. Since most of WSN applications requires
low to moderate bandwidth, a loss in the bandwidth efficiency can be
tolerable, in particular for the unlicensed band applications where
large bandwidth is available. It also found that OOK has a
significant energy saving as compared to the optimized NC-MFSK in
dense WSNs with small values of path-loss exponent. While, for the
indoor environments where the path-loss exponent is large, the
performance difference between OOK and the optimized NC-MFSK
vanishes as the distance between the sensor and sink nodes
increases. In this case, the optimized NC-MFSK is attractive in
use for the band-limited and sparse indoor WSNs.


\end{document}